\documentclass[aps,twocolumn,showpacs]{revtex4-1}
\usepackage{graphicx}
\usepackage{amsmath}
\def\bsigma{\mbox{\boldmath $\sigma$}}
\begin{document}

\title{Electronic states in heterostructures formed by  ultranarrow layers}
\author{F. T. Vasko}
\email{fedirvas@buffalo.edu}
\author{V. V. Mitin}
\affiliation{Department of Electrical Engineering, University at Buffalo, Buffalo, NY 14260-1920, USA}
\date{\today}

\begin{abstract}
Low-energy electronic states in heterosrtuctures formed by ultranarrow layers (single or several monolayers thickness) are studied theoretically. The host material is described within the effective mass approximation and effect of ultranarrow layers is taken into account within the framework of the transfer matrix approach. Using the current conservation requirement and the inversion symmetry of ultranarrow layer, the transfer matrix is evaluated through two phenomenological parameters. The binding energy of localized state, the reflection (transmission) coefficient for the single ultranarrow layer case, and the energy spectrum of superlattice are determined by these parameters. Spectral dependency of absorption due to photoexcitation of electrons from localized states into minibands of superlattice is determined by the ultranarrow layers characteristics. Such a dependency can be used for verification of the transfer matrix and should modify characteristics of optoelactronic devices with ultranarrow layers. Comparision with experimental data shows that the effective mass approach is not valid for description of  ultranarrow layer.
\end{abstract}
\maketitle

\section{Introduction}
Multi-layer heterostructures are widely used in different devices, such as bi- or monopolar heterosrtucture lasers, photodetectors and solar cells, see reviews \cite{1} or \cite{1a}, \cite{2}, and \cite{3}, respectively. Thicknesses of layers in such structures are varied widely, starting from an ultranarrow layer (UNL) of thickness $a$ or $Na$ ($a$ is the monolayer thickness and $N=1,2,\ldots$ is a number of monolayers) for  a single or several monolayer structures, e.g. the case of the quantum dot sheets with ultranarrow wetting layers, see Refs. 5. The case of $N\gg 1$ corresponds to a wide well or a barrier which are described in the framework of the effective mass approximation (EMA) or the $\bf kp$-method supplied by appropriate boundary conditions at heterointerfaces. \cite{5} 
Early in the eighties, similar approaches were used for description of abrupt heterojunctions between different bulk semiconductors, see \cite{6,7} and references in Ref. 8. The electronic properties of short-period superlattices (SL), i. e. heterostructures formed by periodical UNLs, were studied experimentally and numerically based on different  approximations, see Refs. 9 and 10, respectively. These approaches determine the energy spectra of SL only but they are not suitable for consideration of transport and optical phenomena. To the best of our knowledge, an effect of UNLs on these phenomena was not considered in details because the effective mass approximation is not valid over scales $\sim a$. If a heterostructure includes UNL, one should consider it as {\it a new object} which is described by boundary conditions added to the EMA equations at UNL's positions, $z=z_0$. Because heterostructures with UNLs are routinely used in different optoelectronic devices without an investigation of the peculiarities mentioned, it is important and timely to develop an adequate theory of electronic states at UNL and to perform a verification of UNL's parameters.
\begin{figure}[tbp]
\begin{center}
\includegraphics[scale=1.5]{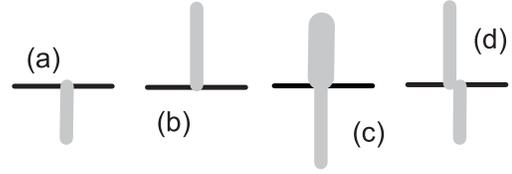}
\end{center}
\addvspace{-1 cm}
\caption{Band diagrams of UNLs enclosed by host semiconductor for: (a) well-like case with $|\tau |<1$, (b) barrier-like case with $\tau >1$ , and (c) combined case with $\tau <-1$, and (d) non-symmetric case. Here $\tau$ is the diagonal element of transfer matrix, see Eq. (5). }
\end{figure}

In this paper we consider an UNL placed at $\{ z_0 \}$, by applying the effective mass approximation (or the $\bf kp$-method) for the host material, at  $z\neq z_0$, and by using boundary conditions at $z\to z_0$ which describe  modifications of envelope functions at UNL. In order to describe low-energy electron states, with energies in the vicinity of the conduction band extremum, in heterosrtuctures formed by UNLs we employ the EMA-approximation in the host material and the boundary conditions at UNLs written through the transfer matrix. The parameters of such a matrix are restricted by the current conservation condition and the boundary conditions are written below for a symmetric UNL, see Figs. 1a - 1c (the case of non-symmetric UNL, shown in Fig. 1d will be considered elsewhere).  Depending on the diagonal element of transfer matrix $\tau$ introduced in Eq. (5), one should separate of UNLs into three categories: (i) a well-like UNL with $|\tau |<1$, (ii) a barrier-like UNL with $\tau >1$, and (iii) a non-conventional UNL with $\tau <-1$. These cases are sketched in Figs. 1a - 1c, respectively, where thicknesses of layers (filled in by grey) are comparable to $a$.  While the cases (i) or (ii) correspond to EMA description of a narrow well or a barrier at $Na\to 0$,  \cite{10} a non-conventional UNL has no simple EMA analogy. In general, it is convenient to characterize the transfer matrix by a dimensionless phase, $\Theta$, and a characteristic wave vector, $K$, see Eqs. (6) and (7) below. These phenomenological parameters determine a character of low-energy states together with the effective mass (or parameters of $\bf kp$-method, if energy is comparable to gap of host semiconductor) and they should be determined from experimental data.

Here we have evaluated the transfer matrix which is written through the parameters $\Theta$ and $K$ (or through characteristic energy $E_K$) and have considered electronic properties of different types of single UNLs as well as SL formed by UNLs. We calculated the binding energy of localized state at a single UNL, which appears to be allowed state for the cases (i) and (iii) and to be forbidden state for the barrier-like case (ii). The energy-dependent reflection (transmission) coefficient is found when considering the scattering of electron on UNL. For the SL formed by UNLs, we analyze the energy spectrum, the density of states, and the absorption coefficient due to photoexcitation of localized electrons under THz or mid-IR excitation. In addition, we demonstrate that these properties are modified essentially under variation of the UNL's characteristics and we discuss possibilities for verification of the transfer matrix parameters.

The consideration below is organized as follows. In Sect. II we describe an UNL in the framework of the transfer matrix approach. The single-layer case is considered in Sect. III and the properties of SL formed by UNL are considered in Sect. IV. Discussion on experimental verification of UNL's parameters is performed in Sect. V. Concluding remarks and list of assumptions used are given in the last section.

\section{Transfer matrix approach}
Within the one-band effective mass approximation, the electronic states in bulk semiconductor with an UNL placed at $z=z_0$ is described by the Schrodinger equation: 
\begin{equation}
\frac{{\hat p_z^2 }}{{2m}}\Psi _z  = E\Psi _z , ~~~~ z \ne z_0  .
\end{equation}
Here $m$ is the effective mass and $\hat p_z$ is the momentum operator. The second-order differential equation (1) should be solved with the boundary conditions which connect the wave functions and their first derivatives at $z_0-0$ and $z_0+0$. Similarly to the consideration of an abrupt heterojunction case, \cite{6,7} we write the connection rules for the column $\Phi_z$ in the form:
\begin{equation}
\Phi _{z_0  + 0}  = \hat T\Phi _{z_0  - 0} , ~~~~ \Phi _z  \equiv \left| {\begin{array}{*{20}c}
   {\Psi _z }  \\
   {d\Psi _z /dz}  \\
\end{array}} \right| .
\end{equation}
The $2\times 2$ transfer matrix $\hat T$ is determined by the current conservation requirement and by the symmetry properties of layer. 

From Eq. (1) it follows that $\widetilde\Phi_z^+\hat\sigma_y\Phi_z$ does not dependent on $z$ if $z\neq z_0$ (here and below $\hat\bsigma$ is the Pauli matrix which connects $\Psi_z$ and its derivative). We impose the current conservation requirement at UNL as $\widetilde\Phi_z^+\hat\sigma_y\Phi_z |_{z_0-0}^{z_0+0}=0$. Since two arbitrary states $\widetilde\Phi_z$ and $\Phi_z$ are considered here, the current conservation gives the condition for the transfer matrix:
\begin{equation}
\hat{T}^+\hat\sigma_y\hat{T}= \hat\sigma_y ,
\end{equation}
see similar evaluation for heterojunction in Ref. 8.
Further, we restrict ourselves by the case of symmetric UNL, when $\Phi_z$ and $\hat\sigma_z\Phi_{-z}$ should be determined by the same equations. As a result one obtains the additional condition:
\begin{equation}
\hat \sigma_z\hat{T}\hat\sigma_z =\hat{T}^{-1} 
\end{equation}
and 4 complex parameters of $\hat T$ should be determined from 8 conditions given by Eqs. (3) and (4). Straightforward calculations give us the transfer matrix
\begin{equation}
\hat T = \left| {\begin{array}{*{20}c}
   \tau  & {\tau_{12} }  \\
   {\tau_{21} } & \tau   \\
\end{array}} \right|, ~~~~ \det \hat T = 1 
\end{equation}
written through real parameters $ \tau$, $\tau_{12}$ and $\tau _{21}$. Since the determinant is fixed here, the matrix $\hat T$ depends on two real parameters which should be considered as phenomenological characteristics of UNL.

At $\left|\tau  \right| <1$, it is convenient to introduce the phase, $\Theta$, and wave vector, $K$, so that the transfer matrix takes the form:
\begin{equation}
\hat T = \left| {\begin{array}{*{20}c}
   {\cos\Theta } & {K^{ - 1} \sin\Theta }  \\
   { - K\sin\Theta } & {\cos\Theta }  \\
\end{array}} \right|
\end{equation}
which depends on two phenomenological parameters, $K>0$ and $0<\Theta <2\pi$.  If $\left| \tau  \right| >1$, one can use similar expressions for $\hat T_ \pm$ which is written through the hyperbolic functions:
\begin{equation}
\hat T_ \pm   = \left| {\begin{array}{*{20}c}
   { \pm \cosh \Theta } & {K^{ - 1} \sinh\Theta }  \\
   {K\sinh\Theta } & { \pm \cosh\Theta }  \\
\end{array}} \right|
\end{equation}
Here the signs $+$ and $-$ correspond to $\tau >1$ and $\tau <-1$, respectively. Once again, we suppose $K>0$ in Eq. (7) because  $\Theta$ can have any sign.

The explicit expressions for $\Theta$ and $K$ can be written for $N\gg 1$, 
when $\hat T$ can be evaluated within the EMA approximation, for the wide layer cases (i) and (ii). For the $N$-monolayer well or barrier of the thickness $Na$ with the band offset $\Delta U$ and the effective mass $m_l$, one obtains the transfer matrix $\hat T$ or $\hat T_+$ given by Eqs. (6) or (7), respectively. The parameters $\Theta$ and $K$ are determined by
\begin{equation}
\Theta =N\sqrt{\frac{m_l}{m}}Ka, ~~~~~ E_K =\frac{(\hbar K)^2}{2m}=\frac{m}{m_l}\Delta U ,
\end{equation}
where the characteristic energy $E_K$ is introduced. For the case of InAs well placed in GaAs \cite{11} one obtains that the energy $E_K$ varies between 1 - 2.8 eV depending on the mismatch stress contributions. The phase $\Theta$ varies between $0.32N$  - $0.42$N, where $N\gg 1$ [for the case (i), $\Theta$ should be reduced to the interval (0, $2\pi$)]. Similar estimates can be obtained for other heterostructures with well- or barrier-like UNL. Notice, that the non-conventional case $\tau <-1$ described by matrix $\hat T_-$ has no analogy with the results for well or barrier described by Eq. (8) within the EMA approximation.

\section{Single-layer case}
Here we consider the single UNL placed at $z_0=0$. The localized state, with energy $-E_0<0$, is described by the wave function
\begin{equation}
\Psi _z  = \Psi _0 \left[ {\theta (z)e^{ - \kappa z}  + \theta ( - z)e^{\kappa z} } \right],
~~~ E_0 =\frac{(\hbar\kappa )^2}{2m} ,
\end{equation}
where $\theta (z)$ is the Heaviside step function, $ E_0$ is the binding energy written through the characteristic size of localization, $\kappa^{-1}$, and $\Psi_0=\sqrt{\kappa}$ is the normalization coefficient. The boundary condition (2) gives us the dispersion relation which determines $\kappa >0$ through the transfer matrix parameters, $\Theta$ and $E_K$. As a result, at $\tau <1$ the binding energy of localized state $E_0$ is given by:
\begin{equation}
E_0 =E_K\left\{ \begin{array}{*{20}c}
 \left(\frac{\sqrt{1+\tan^2\Theta}-1}{\tan\Theta}\right)^2 , & \left|\tau  \right| <1  \\
\left( \frac{\sqrt{1-\tanh^2\Theta}+1}{\tanh\Theta}\right)^2 , & \tau <- 1 \\  \end{array} \right.  
\end{equation}
and there is no localized state for the barrier-like case $\tau >1$.
\begin{figure}[tbp]
\begin{center}
\includegraphics[scale=0.85]{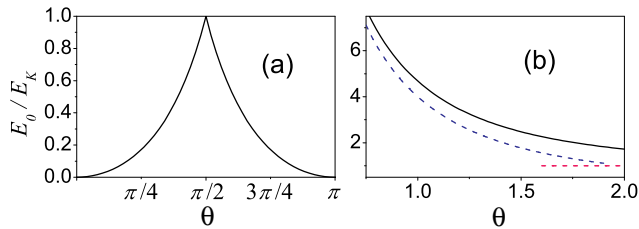}
\end{center}
\addvspace{-0.5 cm}
\caption{(Color online) Dimensionless energy of localized state $E_0/E_K$ versus phase $\Theta$ for the cases of $|\tau |<1$ (a) and $\tau <-1$ (b). Dashed curves in panel (b) correspond to asymptotes at $\Theta\ll 1$ and $\Theta\gg 1$. }
\end{figure}

In Fig. 2 we plot the dependencies of $E_0/E_K$ versus $\Theta$. If $|\tau |<1$ then $E_0/E_K<1$, moreover a deep level with $E_0\to E_K$ is realized at $\Theta\to\pi /2$ and such a dependency is plotted over the interval (0, $\pi$) because it is the periodical function. At $\Theta\leq\pi /4$ or $\pi -\Theta\leq\pi /4$ a shallow state at UNL is realized. For a non-conventional UNL with $\tau\leq -1$, a deep level with $E_0/E_K>1$ is realized and $E_0\to E_K$ if $|\Theta |\gg 1$ while in the opposite case $|\Theta |\to 0$ one obtains the divergent asymptote $E_0/E_K\approx (2/\Theta )^2$. These asymptotes are also shown in Fig. 2b. 
\begin{figure}[tbp]
\begin{center}
\includegraphics[scale=0.85]{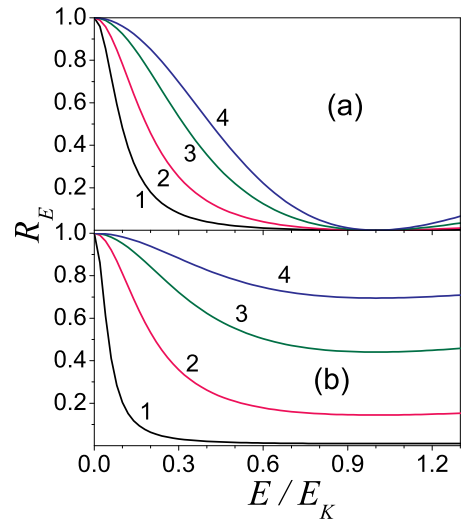}
\end{center}
\addvspace{-0.5 cm}
\caption{(Color online) (a) Reflection coefficient $R_E$ versus dimensionless energy $E/E_K$ for the case $|\tau |<1$ at $\Theta =\pi /16$ (1), $\pi /8$ (2), $\pi /4$ (3), and $\pi /2$ (4). (b) The same for the case $|\tau |>1$ at $\Theta =0.1$ (1), 0.4 (2), 0.8 (3), and 1.2 (4).}
\end{figure}

The scattering problem, for an electron with energy $E>0$ propagating from the left, is described by the wave function
\begin{equation}
\Psi_z =\theta (-z)\left(\Psi_i e^{ikz} +\Psi_r e^{-ikz}\right) +\theta (z)
\Psi_t e^{ikz} ,
\end{equation}
where $\Psi_i$, $\Psi_r$, and $\Psi_t$ are the amplitudes of incident, reflected, and transmitted waves. The wave vector $k$ is connected with the energy $E$ by the standard relation $E = (\hbar k)^2 /2m$. The reflected and transmitted amplitudes, $\Psi_r$, and $\Psi_t$, are expressed through $\Psi_i$ from the boundary condition (2). Further, we introduce the flows $J_\gamma   = \left| {\Psi _\gamma  } \right|^2 \hbar k/m$, which correspond to the incident ($\gamma   =i$), reflected ($\gamma   =r$), and transmitted ($\gamma   =t$) waves, and calculate the reflection and transmission coefficients according to $R_E=J_r/J_i$ and $T_E=J_t/J_i$. Since the particle conservation law, $R_E+T_E=1$, only the  reflection coefficient is considered below:
\begin{eqnarray}
R_E  = \left[ {1 + F(\Theta ,k/K)} \right]^{ - 1}  ,  \nonumber  \\
F(\Theta ,z) = \left\{ {\begin{array}{*{20}c}
 {\left[ {\frac{{2z}}{{\sin\Theta (1 - z^2 )}}} \right]^2 ,} & {\left| \tau  \right| < 1}  \\
 {\left[ {\frac{{2z}}{{\sinh\Theta (1 + z^2 )}}} \right]^2 ,} & {|\tau |>1}  \\
\end{array}} \right. .
\end{eqnarray}
Here $F(\Theta ,z) =F(-\Theta ,z) $ and $R_E$ is determined by positive $\Theta$; the interval $0<\Theta<\pi /4$ is enough in order to describe the case of UNL with $|\tau |<1$.

Fig. 3 shows the reflection coefficient $R_E$ versus $E/E_K=(k/K)^2$. Since $F(\Theta ,0)=0$, there is no tunneling of low-energy electrons through any UNL because $R_{E\to 0}\to 1$. If energy increases, $R_E$ decrease up to $E=E_K$ and, once again, $R_{E}\to 1$ if $E\gg E_K$. But in the region $E\sim E_K$ behavior of $R_E$ is different for $|\tau |<1$ and $|\tau |>1$: the well-like case $R_{E_K}=0$, while for $|\tau |<1$ the reflection coefficient approaches constant, compare Figs. 3a and 3b [there is no difference between the cases (ii) and (iii) in panel (b)].

\section{Superlattice}
We turn now to the consideration of electronic states in SL of the period $l$ formed by the UNLs placed at $z_0  \to \left\{rl\right\}, ~r = 0, \pm 1, \cdots$. Based on the boundary conditions (2) at $z=rl$, below we consider the miniband energy spectrum and calculate the absorption coefficient of doped SL under photoexcitation of localized states.

\subsection{Miniband spectrum}
The eigenvalue problem for the periodic system of UNLs is solved introducing a quasimomentum $0<p_\bot <2\pi\hbar /l$ and using the Bloch's theorem, $\Psi_{p_\bot z}=\exp (ip_\bot l/\hbar)\Psi_{p_\bot z-l}$. The wave function takes form 
\begin{equation}
\Psi_{p_\bot z}=\frac{N_{p_\bot}}{\sqrt{l}}\left[ e^{ikz}+R(p_\bot ,k)e^{-ikz}\right]  ,
~~~~ 0 < z < l ,
\end{equation}
where $N_{p_\bot}$ is the normalization factor. By analogy with the general consideration, \cite{10} the coefficient $R(p_\bot ,k)$ and the dispersion relation between $p_\bot$ and $k$ are determined from the periodicity requrement and the boundary conditions (2) at $z=0,~l$. For the UNL with $|\tau |<1$ we use the transfer matrix (6) and the factor $R(p_\bot ,k)$ takes form
\begin{equation}
R(p_\bot ,k)=\frac{e^{ip_\bot l/\hbar}-\left[\cos\Theta +(ik/K)\sin\Theta e^{ikl}\right]}{e^{ip_\bot l/\hbar}-\left[\cos\Theta -(ik/K)\sin\Theta e^{-ikl}\right]} ,
\end{equation}
while the dispersion relation is written as follows:
\begin{equation}
\cos\frac{p_\bot l}{\hbar}=\cos\Theta\cos kl -\frac{1}{2}\left(\frac{K}{k}+\frac{k}{K}\right)\sin\Theta\sin kl .
\end{equation}
If $|\tau |>1$ one uses the transfer matrix (7) and the factors $R_\pm (p_\bot ,k)$ should be written similarly to Eq. (14) but through the hyperbolic functions, $\pm\cosh\Theta$ and $\sinh\Theta$ instead of $\cos\Theta$ and $\sin\Theta$. The corresponding dispersion relation is given by
\begin{eqnarray}
\cos\frac{p_\bot l}{\hbar}=\pm\cosh\Theta\cos kl \nonumber  \\
+\frac{1}{2}\left(\frac{K}{k}-\frac{k}{K}\right)\sinh\Theta\sin kl .
\end{eqnarray}
As in Eq. (7), here and in $R_\pm (p_\bot ,k)$ the signs $+$ and $-$ correspond to $\tau >1$ and $\tau <-1$, respectively. The last case has no analogy with the standard dispersion relations obtained in the EMA approximation, while Eq. (15) and Eq. (16) with $+\cosh\Theta$ are similar to the results for well or barrier of finite width. \cite{10}

The contour plots of the right-hand sides of Eqs. (15) and (16) versus $\Theta$ and dimensionless energy $E/\sqrt{E_K\varepsilon_l}$, [here  $\varepsilon_l =(\hbar /l)^2/2m$ is a small energy corresponding to SL of period $l$] are presented in Fig. 4.  Here the gap regions, which are above +1 and below $-1$, are shaded by grey and the dashed curves are plotted at the middle of miniband energies, when $\cos p_\bot l/\hbar =0$. For the well-like case, $|\tau |<1$, the weakly-coupled SL is realized if $\Theta\to 0,~\pi ,~2\pi$ and far from these points the  tight-binding coupling regime takes place, see Figs. 4a and 4b. With increasing of the dimensionless parameter $\sqrt[4]{E_K/\varepsilon_l}$, a number of minibands increases and a weakly-coupled regime is under more rigid conditions, compare Figs. 4a and 4b. If $|\tau |>1$ (i.e. SL is formed by barrier-like or non-conventional UNLs), the tight-binding regime of coupling takes place if $\Theta >0.5$ and a weakly-coupled SL is realized at $\Theta\to 0$. Transformations between these regimes are different if $\tau >1$ or $\tau <-1$, compare Figs. 4c and 4d.
\begin{figure}[tbp]
\begin{center}
\includegraphics[scale=0.2]{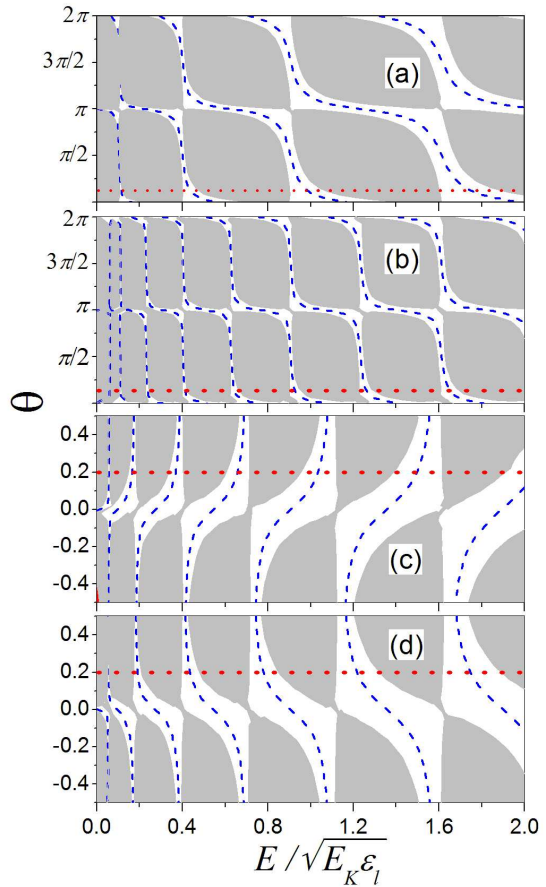} 
\end{center}
\addvspace{-0.5 cm}
\caption{ (Color online)  (a) Contour plot of the right-hand side of Eq. (15) versus $\Theta$ and dimensionless energy $E/\sqrt{E_K\varepsilon_l}$ at $\sqrt[4]{E_K/\varepsilon_l}=$10. (b) The same as in panel  (a)  at $\sqrt[4]{E_K/\varepsilon_l}=$20. (c) The same as in panel (a) for dispersion equation (16)  at $\tau >1$ and $\sqrt[4]{E_K/\varepsilon_l}=$15. (d) The same as in panel (c) at $\tau <-1$. Gap regions are shaded, dashed curves correspond to zero level, and dotted lines correspond to cross-sections shown in Fig. 5 below. }
\end{figure}

The dispersion laws $\varepsilon_{rp_\bot}=(\hbar k_{rp_\bot})^2/2m$, which are written through $k_{rp_\bot}$ determined by the dispersion equations (15) or (16) depending on UNL parameters, are shown in Fig. 5 at $\Theta$ marked in each panel of Fig. 4. Energy scale of gaps and allowed bands is determined by the characteristic energy $\sqrt{E_K\varepsilon_l}$ which is between 70 - 25 meV for $l=$10 nm - 50 nm if $E_K$ is taken as 1 eV. The dispersion laws plotted in Fig. 5 are close to cosine or sine dependencies.
\begin{figure}[tbp]
\begin{center}
\includegraphics[scale=1]{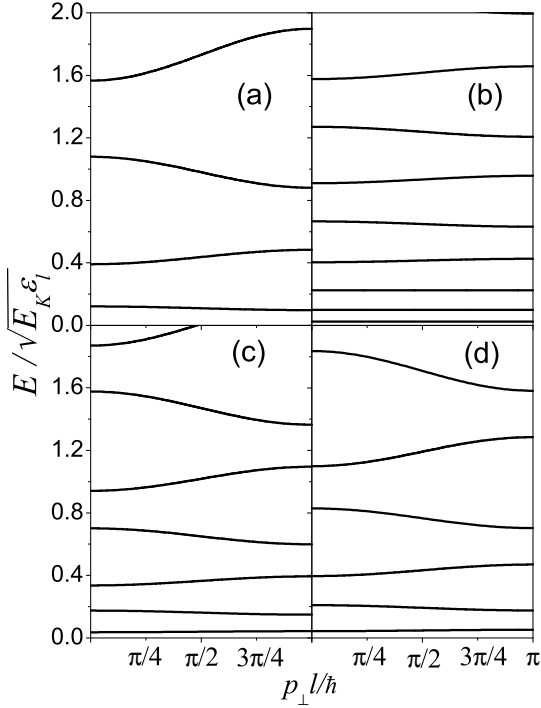}
\end{center}
\addvspace{-0.5 cm}
\caption{Miniband energy spectra $\varepsilon_{rp_\bot}$ for the parameters used in Figs. 4a-4d at $\Theta =\pi /8$ (a, b) and 0.2 (c,d). }
\end{figure}

Farther, we consider the density of states which is introduced by the standard formula: $\rho_E =(2/L^3)\sum\nolimits_\delta\delta (E-\varepsilon_\delta )$ where $\varepsilon_\delta\to\varepsilon_{rp_\bot}+\varepsilon_p$. Here we separated the in-plane kinetic energy, $\varepsilon_p =p^2/2m$, which corresponds to the 2D momentum $\bf p$, and the superlattice contribution, $\varepsilon_{rp_\bot}$ plotted in Fig. 5. Integration over $\bf p$ gives the 2D density of states, $\rho_{2D}$, and $\rho_E$ appears to be written through the integrals taken over step function, $\theta (z)$:
\begin{equation}
\rho_E =\rho_{2D}\sum\limits_r\int\limits_0^{2\pi\hbar /l}\frac{dp_\bot}{2\pi \hbar}\theta\left( E-\varepsilon_{rp_\bot}\right) .
\end{equation}
If $E$ belongs to $\bar r$th gap, the $\theta$-functions for $r\leq\bar r$ should be replaced by unit and $\rho_E$ is constant. In $\bar r$th miniband (below $r$th gap), the integral over $p_\bot$ should be taken over the interval $(0,p_E)$ where $p_E$ is found as a root of the equation $E=\varepsilon_{\bar rp_E}$. As a result, the density of states takes the form:
\begin{equation}
\rho_E =\frac{\rho_{2D}}{l}\left\{ \begin{array}{*{20}c}
\bar r, & E \subset  \bar r{\rm th ~ gap}  \\
\bar r-1+\frac{p_E l}{2\pi \hbar} , & E \subset  \bar r {\rm th ~ miniband} \\
\end{array} \right.
\end{equation}
and a ladder-like shape of $\rho_E$ is determined by the gap-induced steps with miniband contributions between them.

In Fig. 6 we plot the density of states for the same parameters as in Fig. 5. Here the thick straight lines correspond to gap contributions with the miniband contributions between them which are similar to arccosine dependencies. One can see that a number of minibands increases with the parameter $\sqrt[4]{E_K/\varepsilon_l}$. An approach to the square-root dependence, corresponding to the bulk density of states, takes place at $E/\sqrt{E_K\varepsilon_l}>$3. Since $\rho_E$ is connected directly to the shape of interband optical spectra, see Ref. 6b, the step-like dependencies over the interval $E/\sqrt{E_K\varepsilon_l}\leq $3 permit one to extract UNL parameters which determine the bandstructure of SL. Effect of the above-barrier states on PLE spectra in wide InGaAs/GaAs structure was measured and calculated within the EMA approach in Ref. 12. 
\begin{figure}[tbp]
\begin{center}
\includegraphics[scale=1]{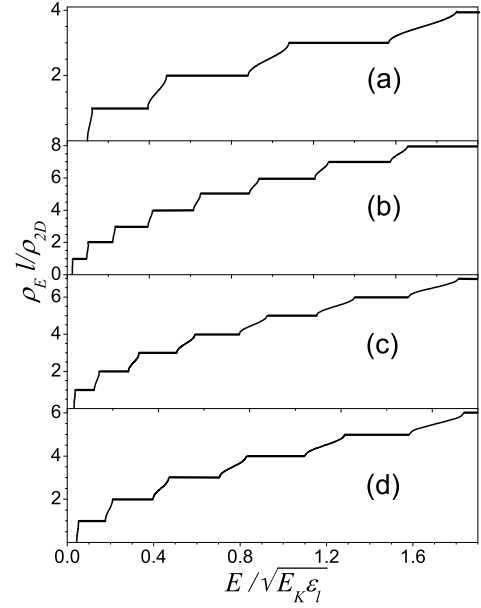}
\end{center}
\addvspace{-0.5 cm}
\caption{Normalized density of states for the parameters used in panels (a-d) of Figs. 4 and 5. }
\end{figure}

\subsection{Absorption coefficient}
Using the solutions obtained, we consider below the process of photoexcitation of electrons localized at levels of energy $-E_0$ into minibands caused by the radiation polarized along SL axis. Such absorption takes place in heavily doped SL formed by the well-like or non-conventional UNLs; we do not consider here the case (ii) which is similar to an ordinary SL, see Ref. 13. The absorption coefficient $\alpha_\omega$ is determined by the general Kubo formula as follows:
\begin{eqnarray}
\alpha_\omega =\frac{8(\pi e)^2}{\sqrt\epsilon c\omega L^3}\sum\limits_{\delta \delta '}\left[ f(\varepsilon_\delta )-f(\varepsilon_\delta +\hbar\omega )\right]  \nonumber \\
\times\left| (\delta |\hat{v}_z |\delta ')\right|^2\delta\left(\varepsilon_\delta  
-\varepsilon_{\delta '}+\hbar\omega\right) ,
\end{eqnarray}
where $\epsilon$ is the dielectric permittivity of the host semiconductor, $ L^3$ stands for the normalization volume, and the matrix element $\left| (\delta |\hat{v}_z |\delta ')\right|^2$ corresponds to transitions between $\delta$- and $\delta '$-states of energies $\varepsilon_\delta$ and $\varepsilon_{\delta '}$.  We use the equilibrium distribution $f(\varepsilon_\delta )$ and take into account $\varepsilon_\delta\to\varepsilon_p -E_0$ because only localized states are  populated. Since transitions are vertical, the energy conservation law and the matrix element do not depend on $\bf p$ and the 2D concentration $n_{2D}=(2/L^2)\sum\nolimits_{\bf p} f(\varepsilon_p -E_0)$ appears in (19) if $\hbar\omega > E_0$. In addition, the matrix element $M_{rp_\bot}=\left| (0|\hat v_z |rp_ \bot  )\right|^2$ is the same for any UNL and (19) may be transformed into the sum over different minibands
\begin{equation}
\alpha_\omega =\frac{2\pi e^2 n_{2D}}{\sqrt\epsilon c\hbar\omega}  
\sum\limits_r\int\limits_0^{2\pi\hbar /l}dp_\bot M_{rp_\bot}
\delta\left({\hbar\Delta\omega -\varepsilon_{rp_\bot} }\right) 
\end{equation}
written through the frequency detuning, $\Delta\omega =\omega -E_0 /\hbar$.

The transparency regions take place if $\hbar\Delta\omega$ belongs to any gap. If $\hbar\Delta\omega$ is brought into the $\bar r$th miniband, the corresponding absorption coefficient takes the form 
\begin{equation}
\alpha_\omega ^{(\bar r)}  = \frac{{2\pi e^2 n_{2D} }}{{\sqrt \epsilon  c\hbar \omega }}\frac{M_{\bar rp_{\Delta \omega} }}{|d\varepsilon_{\bar rp_\bot } /dp_\bot  |_{p_{\Delta \omega } }} ,
\end{equation}
where $p_{\Delta\omega}$ should be determined from the equation $\hbar\Delta\omega =\varepsilon_{rp_{\Delta\omega}}$. Under the condition $\kappa l\gg 1$, the matrix element $M_{rp_\bot}$ is written through the wave functions (9) and (14) in the form:
\begin{eqnarray}
M_{rp_\bot}=\left( {\frac{\hbar }{m}} \right)^2 \frac{\kappa }{l}\left| {N_{p_ \bot  } } \right|^2
~~~  \\
\times\left| {1 + e^{ikl}  + R\left( {p_ \bot  ,k} \right)\left( {1 + e^{ - ikl} } \right)} \right|^2 
\nonumber
\end{eqnarray}
for the UNL with $|\tau |<1$. If $|\tau |>1$ one should use $R_{\pm}\left( {p_ \bot  ,k} \right)$ in Eq. (22). Notice, that $M_{rp_\bot}$ and the velocity $d\varepsilon _{\bar rp_ \bot  } /dp_ \bot$ vanish at the edges of minibands, so that jumps of absorption are possible at edges of absorption regions. Thus,
spectral dependencies of $\alpha _\omega^{(\bar r)}$ appear to be strongly dependent on the transfer matrix parameters and they can be extracted from these data. 
\begin{figure}[tbp]
\begin{center}
\includegraphics[scale=1]{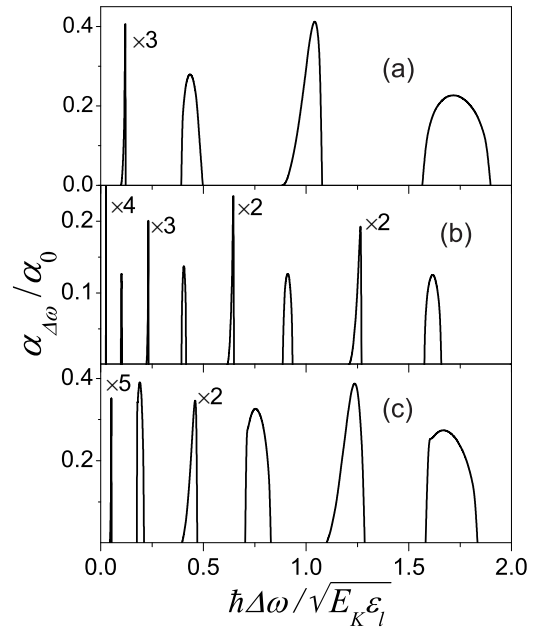}
\end{center}
\addvspace{-0.5 cm}
\caption{(a) Spectral dependencies of dimensionless absorption coefficient $\alpha_{\Delta\omega} /\alpha_0$ for $\Theta =\pi /12$ at the same conditions as in Figs. 4a - 6a. (b) The same as in panel (a) for the conditions of Figs. 4b - 6b. (c) The same as in panels (a,b) for $\Theta =$0.1 and the conditions of Figs. 4d - 6d. }
\end{figure}

The shapes of absorption peaks are determined by Eqs. (21) and (22)  and by the dispersion relations $\varepsilon _{\bar rp_ \bot  }$, as it is shown in Fig. 7 for lower peaks. Here we plotted the dimensionless spectral dependencies $\alpha_{\Delta\omega} /\alpha_0$ where 
\begin{equation}
\alpha _0  = \frac{8\pi e^2}{\sqrt \epsilon  c}n_{2D} \sqrt {\frac{2}{mE_0}} 
\end{equation}
is the characteristic absorption. For the case of IR absorption of SL with concentration $n_{2D}=5\times 10^{11}$ cm$^{-2}$ and $E_0\sim$0.1 eV,\cite{12a} one obtains $\alpha_0\simeq 1.25\times 10^4$ cm$^{-1}$ if $\sqrt{\epsilon}\simeq$3.3. The positions and widths of absorption peaks correspond to the miniband energy spectra shown in Figs. 4 and 5.
There is an essential difference between odd and even peaks due to the inversion symmetry of the eigenstate problem at edges of minibands, when $p_\bot l/\hbar =0,~2\pi$. The odd peaks are non-symmetric ones, they are higher (see the scaling factors in Fig. 7) and more narrow in comparison to the even peaks which are more symmetric and wide. Since the condition $\kappa l>1$ is satisfied at $l\geq$10 nm, a visible ($>$30\%) absorption takes place for 20-layer periodic structure and the mid-IR measurements permit a direct verification of the SL parameters.

\section{Experimental verification}
Next, we briefly discuss possibilities for experimental verification or for {\it ab initio} calculations of the UNL's parameters. It is not a simple task because the UNL's parameters affect on response of heterostructure in a complicated way and one needs to design an appropriate structure and to perform special measurements. Also, any first-principle calculations of the UNL's parameters should include both UNL and host material, so that it is necessary to perform calculations on atomic level for a long period SL in order to compare with the results of Sect. IV written through the UNL's parameters. Any method used for short-period SL, see \cite{8} and Refs. therein, can be applied providing a generalization for the case of a long host material. Similar calculations for InAs UNL embedded by GaAs together with the measurements of capacitance-voltage characteristics were performed in Ref. 15 and the binding energy of localized state $E_0$ appears to be $\sim$60 meV. This value is about 3 times smaller than EMA estimate of $E_0$ given by Eq. (8) and Fig. 2a. Thus, the EMA-approach is not valid for description of the InAs UNL in GaAs host matrix but a complete determination of parameters $\Theta$ and $E_K$ is not possible due to luck of data in Ref. 15.

Below we list a set of papers, where UNLs based on different heteropairs were realized, and discuss possible measurement schemes in order to verify UNL's parameters. Different optoelectronic devices contain the quantum dot sheets with an ultranarrow wetting layer \cite{1,1a,2,3} and influence of these layers on electronic states is described by the results obtained here if one neglects the quantum dots effect. UNLs with $N\sim$5 are widely used in the quantum cascade lasers, \cite{12c} where each period contains about 10 layers of different thickness, so that a verification of UNL's parameters is difficult.
At least several measurements of InAs UNLs placed into GaAs or AlGaAs matrix \cite{13} as well as results for UNLs in A$_{IV}$B$_{VI}$ \cite{14} and in A$_{II}$B$_{VI}$ \cite{15} heterostructures were published. But a verification of UNL's parameters was not performed in these papers and a special investigation for each system is necessary. 

All the results obtained in Sect. IV depend strongly on $\Theta$ and $E_K$, which determine the transfer matrices (6) or (7), and these parameters can be verified from the spectroscopical measurements. Indeed, the results of Sect. IVB demonstrate that mid-IR (or THz, depending on parameters in question) absorption is strongly varied depending on transfer matrix parameters. The near-IR interband transitions in SL depend essentially on the density of states considered in Sect. IVA for different UNL's parameters. Variations of the near-IR spectra from the bulk case permit one to verify $\Theta$ and $E_K$ using the electromodulation spectroscopy or the PLE measurements. Beside this, the transfer matrix parameters can be extracted from the transport data but in addition to UNL's parameters, some other characteristics, e. g. scattering rates, should be involved under interpretation of these data.

\section{Conclusions}
In summary, we have re-examined the theory of electronic states in the heterostructures formed by ultranarrow layers taking into account that the EMA approximation can not be applied for description of UNLs. Our consideration is based on the transfer matrix method which was generalized for the case of UNL. It is found that the three types of UNL are possible depending on a value of diagonal element in the transfer matrix (5). The cases $|\tau |<1$ or $\tau >1$ are similar to narrow well or barrier while there is no simple analogy with a wide layer for the non-conventional UNL with $\tau <-1$. The localized level appears at single UNL with $\tau <1$. The energy-dependent reflection coefficient of a single UNL as well as the energy spectrum of SL formed by a periodical array of UNLs are different for the cases $|\tau |<1$ and $\tau >1$. The spectral dependencies of absorption are analyzed for the case of photoexcitation of localized electrons into SL's minibands. The EMA approach fails to explain the results of Ref. 15 but there is a luck of experimental data for verification of UNL's parameters.

Let us discuss now the main assumptions applied to the consideration performed. Using the single-band Hamiltonian in Eq. (1) we suppose that electron energy $E$ is small in comparison with the gap of the host semiconductor. In order to consider the high-energy states, one needs to use {\bf kp}-Hamiltonian and more complicated boundary conditions, see similar consideration of abrupt heterojunction in Refs. 8 and 21. In addition, the transfer matrix approach is valid for description of UNL of width $Na$ if $Na$ is less than electron wavelength, i. e. $(\hbar /Na)^2/2m\gg E$. We restrict ourselves by the model of symmetric UNL based on the condition (4). It should be mentioned that the EMA-based estimates given by Eq. (8) are not valid for $N\sim 1$ at least for InAs UNLs, see discussion in Ref. 20. A more complicated consideration is necessary in order to take into account a non-symmetry of UNL; this case will be considered elsewhere. Also, we consider ideal UNL neglecting an in-plane scattering processes or inhomogeneous broadening; this approximation is valid if $E$ exceeds a typical broadening energy. A possible segregation of UNLs (some experimental data for InAs see in \cite{17}) will not be essential because a short-scale lateral inhomogeneities is realized. 

To conclude, we believe that the results obtained will stimulate a verification of phenomenological parameters describing the electronic properties of UNLs using the mid-IR spectroscopy when the valence band states are not essential. These data should be important for applications of heterostructures with UNLs in different devices.


\begin{thebibliography}{99}
\bibitem{1} 
{\it Quantum Well Lasers} Ed. by P. S. Zory, (Academic Press, San Diego, 1993).

\bibitem{1a}
C. Gmachl, F. Capasso, D. L. Sivco, and A. Y. Cho, Rep. on Progr. in Phys., 
{\bf 64}, 1533 (2001).

\bibitem{2} 
A. Rogalski, J. Appl. Phys. {\bf 93}, 4355 (2003); J. N. Pan and C. G. Fonstad, Material Sci. and Engineering {\bf 28}, 65 (2000).

\bibitem{3} 
V. Avrutin, N. Izyumskaya, and H. Morkoc, Superlattices and Microstructures,  
{\bf 49}, 337 (2011); G. F. Brown and J. Q. Wu, Laser and Photonics Reviews, 
{\bf 3}, 394 (2009).

\bibitem{4} 
D. Bimberg, M. Grundmann, N. N. Ledentsov, {\it Quantum Dot Heterostructures} (J. Wiley and Sons, New York, 1999); I. L. Krestnikov, N. N. Ledentsov, A. Hoffmann, and D. Bimberg, Physica Status Solidi A, {\bf 183}, 207 (2001). 

\bibitem{5}
G. Bastard, {\it Wave Mechanics Applied to Semiconductor Heterostructures} (Editions de Physique, Paris, 1988); F. T. Vasko and A. Kuznetsov, {\it Electronic States and Optical Transitions in Semiconductor Heterostructures} (Springer, New York, 1998).

\bibitem{6} 
T. Ando and S. Mori, Surf. Sci. {\bf 113}, 124 (1982); I.M. Sokolov, Zh. Eksp. Teor. Fiz. 89, 556 (1985) [Sov. Phys. JETP {\bf 62}, 317 (1985)]; B. Laikhtman, Phys. Rev. B {\bf 46}, 4769 (1992).

\bibitem{7} 
I. V. Tokatly, A. G. Tsibizov, and A. A. Gorbatsevich, Phys. Rev. B, {\bf 65}, 165328 (2002).

\bibitem{8} 
D. Donetsky, S. P. Svensson, L. E. Vorobjev, and G. Belenky, Appl. Phys. Lett.  {\bf 95}, 212104 (2009); H. Li, S. Katz, G. Boehm, and M.-C. Amann, Appl. Phys. Lett. {\bf 98}, 131113 (2011).

\bibitem{9} 
H. Schmidt, R. Pickenhain, and G. Bohm, Phys. Rev. B, {\bf 65}, 045323 (2002).

\bibitem{10} 
M. Herman, {\it Semiconductor Superlattices} (Academie-Verlag, Berlin, 1986);
V. Mitin, D. Sementsov and N. Vagidov, {\it Quantum Mechanics for Nanostructures} (Cambridge University Press,  Cambridge 2010).

\bibitem{11} 
T. Worren, K. B. Ozanyan, O. Hunderi, and F. Martelli, Phys. Rev. B {\bf 58}, 3977 (1998).

\bibitem{12} 
A. Ya. Shik, Sov. Phys. Semiconductors, {\bf 8}, 1841 (1974); {\it ibid.} {\bf 6}, 1110 (1972); A. Kastalsky, T. Duffield, S. J. Allen, and J. Harbison, 
Appl. Phys. Lett. {\bf 52}, 1320 (1988).

\bibitem{12a}
This value is taken in agreement with experimental data of Ref. 15.

\bibitem{12b}
R. Pickenhain, H. Schmidt, and V. Gottschalch, J. Appl. Phys. {\bf 88}, 948 (2000).

\bibitem{12c}
O. Cathabard, R. Teissier, J. Devenson, J. C. Moreno, and A. N. Baranov, Appl. Phys. Lett., {\bf 96}, 141110 (2010);
D. G. Revin, J. P. Commin, S. Y. Zhang, A. B. Krysa, K. Kennedy, and J. W. Cockburn, IEEE J. of Quant. Electr. {\bf 17}, 1417 (2011).

\bibitem{13}
D. Guimard, R. Morihara, D. Bordel, K. Tanabe, Y. Wakayama, M. Nishioka, and Y. Arakawa, Appl. Phys. Lett. {\bf 96}, 203507 (2010);
M. Di Ventra and K. A. Mader, Phys. Rev. B, {\bf 55}, 13148 (1997)

\bibitem{14}
J. D. Jeffers, K. Namjou, Z. Cai, P. J. McCann, and L. Olona, Appl. Phys. Lett. {\bf 99}, 041903 (2011).

\bibitem{15}
S. V. Ivanov, A. A. Toropov, T. V. Shubina, S. V. Sorokin, A. V. Lebedev, I. V. Sedova, P. S. Kopev, G. R. Pozina, J. P. Bergman, and B. Monemar, J. Appl. Phys. {\bf 83}, 3168 (1998).

\bibitem{16}
P. Paki, R. Leonellia, L. Isnard, and R. A. Masut, Appl. Phys. Lett. {\bf 74}, 1445 (1999); V. Albe and L. J. Lewis, Physica B {\bf 301}, 233 (2001).

\bibitem{16a}
M. V. Kisin, B. L. Gelmont, S. Luryi, Phys. Rev. B, {\bf 58}, 4605 (1998).

\bibitem{17}
S. Martini, J. E. Manzoli, and A. A. Quivy, J. Vac. Sci. Technol. B {\bf 28}, 277 (2010); R. Ares, C. A. Tran, and S. P. Watkins,  Appl. Phys. Lett. {\bf 67}, 1576 (1995).

\end{thebibliography}
\end{document}